\documentclass{aa}

\usepackage{graphicx}
\usepackage{txfonts}
\usepackage{amsmath}
\usepackage{xcolor}
\usepackage{soul}
\usepackage{upgreek}
\usepackage{comment}
\usepackage{url}
\usepackage{hyperref}
\usepackage{orcidlink}
\usepackage{booktabs}  % \toprule, \midrule, \bottomrule, \cmidrule
\usepackage{multirow}  % \multirow
\usepackage{graphicx}  
\usepackage{rotating}
\usepackage{xcolor}

\graphicspath{{./}{}}

\begin{document}

   \title{Testing Offsets Between Cluster-Scale Halos and BCGs in Strong Lensing Models Using the Jackknife Method}

   \titlerunning{Cluster-scale Halo Positional Freedom in Strong Lensing}
   \authorrunning{Cha et al.}

   \author{Sangjun Cha\inst{\ref{in:ASIAA},\ref{in:Yonsei}}\corrauth{scha@asiaa.sinica.edu.tw}%\fnmsep\thanks{EACOA Fellow}\,
   \orcidlink{0000-0001-7148-6915}
           \and Marceau Limousin\inst{\ref{in:CNRS}},\orcidlink{0000-0001-6636-4999}
           \and M. James Jee\inst{\ref{in:Yonsei},\ref{in:Davis}},\orcidlink{0000-0002-5751-3697}
           }

   \institute{Academia Sinica Institute of Astronomy and Astrophysics (ASIAA), No. 1, Section 4, Roosevelt Road, Taipei 106319, Taiwan \label{in:ASIAA}
         \and Department of Astronomy, Yonsei University, 50 Yonsei-ro, Seoul 03722, Korea \label{in:Yonsei}
         \and Aix Marseille Univ, CNRS, CNES, LAM, Marseille, France \label{in:CNRS}         
         \and Department of Physics, University of California, Davis, One Shields Avenue, Davis, CA 95616, USA \label{in:Davis}
              }

   \date{Received XXX; accepted XXX}

   \abstract
   {
   Offsets between cluster-scale dark matter halos and brightest cluster galaxies (BCGs) inferred from strong lensing (SL) models have been used to investigate cluster dynamics and dark matter physics. However, it is necessary to test whether these offsets are required by SL data themselves. We investigate whether allowing these offsets improves predictive accuracy and examine the effects on magnification, time-delay predictions, and critical curve positions. We evaluate lens models of Abell 370, RX J1347.5-1145, and MACS J0416.1-2403 using the jackknife method, which assesses predictions for source systems excluded from the lens reconstruction. Models are constructed with {\tt MrMARTIAN}, which combines analytic mass profiles with a regularized grid that represents mass structure not captured by the profiles. We compare fixed and free halo position models across regularization weights. The root mean square values averaged over excluded systems agree within bootstrap uncertainties, while the median paired differences remain close to zero across all three clusters and tested regularization weights. These results indicate comparable predictive accuracy between the fixed and free models. The total mass distributions remain broadly similar, although the magnitudes of fitted halo offsets vary with regularization, showing that the mass decomposition is not unique. For these three clusters, offsets between BCGs and cluster-scale halos in the lens models are not required by the SL data to improve predictive accuracy within the MrMARTIAN framework. The mass decomposition between the regularized grid and cluster-scale halos is not uniquely determined. These findings limit physical interpretations based on fitted halo positions alone. The model dependence of magnification and time-delay estimation can also affect the inferred intrinsic properties of high-redshift sources or the Hubble constant.
   }


   \maketitle

\section{Introduction}
Strong gravitational lensing (SL) by galaxy clusters provides a direct probe of mass distributions without assuming dynamical equilibrium and is used in a wide range of studies. For example, time delays between multiple images of lensed supernovae, combined with lens model predictions, give constraints on the Hubble constant $H_0$ \citep[e.g.,][]{2023Sci...380.1322K,2025ApJ...979...13P,2026A&A...708A.291S}. High magnification around critical curves enables observations of individual stars at cosmological distances and studies of their intrinsic properties \citep[e.g.,][]{2018PhRvD..97b3518O,2018NatAs...2..334K,2019A&A...625A..84D,2025NatAs...9..428F,2025A&A...699A.299M,2025A&A...704A.309L,2026arXiv260422702P}. The spatial distribution of lensed stellar transients has also been used to investigate the nature of dark matter (DM) \citep[e.g.,][]{2025ApJ...978L...5B}. Current JWST programs, including SLICE \citep{2026ApJ..1001...60C} or VENUS \citep{2025jwst.prop.6882F}, are expanding the samples of galaxy clusters with deep, high-resolution near-infrared imaging. These observations reveal new SL multiple images and resolve structures in lensed sources, providing additional constraints for cluster SL modeling. Since the aforementioned lensing-based studies rely on lens model predictions, the reliability of the reconstructed lens models is important.

Cluster SL models sometimes show large offsets between the fitted centers of cluster-scale DM halos and the brightest cluster galaxies (BCGs). Theoretical and numerical studies predict that group- and cluster-scale DM halos should be associated with bright cluster galaxies \citep[e.g.,][]{2024OJAp....7E..65R}, yet some SL models \citep[e.g.,][]{2019MNRAS.485.3738L, 2023MNRAS.524.2883N} report offsets larger than those allowed by self-interacting DM (SIDM) scenarios \citep[e.g.,][]{2017MNRAS.469.1414K} or include DM halos without luminous counterparts (light-unaffiliated mass clumps, or LUMCs). 
Such offsets can be regarded as additional freedom in lens modeling, but some studies have tried to interpret them physically. 
Merging clusters have been used to study cluster dynamics and constrain DM self-interactions, for example by comparing the distributions of DM and cluster member galaxies \citep[e.g.,][]{2008ApJ...679.1173R, 2015Sci...347.1462H,2025ApJ...987L..15C} or the positions of DM halos and radio relics \citep[][]{2026ApJ..1007...56J}. Offsets between BCGs and the centers of their host halos have also been compared with simulations to constrain the SIDM cross section \citep{2019MNRAS.488.1572H}.
However, the physical interpretation of offsets inferred from SL modeling remains unclear. A fitted displacement may reflect a physical separation, limitations in the adopted mass representation, or both. Before interpreting these offsets in terms of cluster dynamics or DM physics, it is necessary to test whether they are required by the SL data themselves.

This question has motivated studies that consider the relation between mass and light in cluster lens models. \citet{2022A&A...664A..90L} revisited three parametric models featuring LUMCs and were able to reproduce the SL constraints equally well with models in which each group- or cluster-scale DM halo is associated with a bright cluster galaxy. \citet{2025A&A...703A..10L} revisited the JWST based mass model of MACS J0416.1-2403, for which previous parametric SL studies presented models featuring one cluster-scale LUMC. They were able to remove this LUMC, presenting a competitive mass model in which each cluster-scale DM halo is associated with a luminous counterpart. By contrast, \citet{2025A&A...693A..33L} were unable to propose a parametric mass model in which DM is traced by light in Abell 370. To reach a sub-arcsecond image plane root mean square (RMS), a cluster-scale LUMC had to be introduced in the modeling, as also found in earlier works \citep{2019MNRAS.485.3738L,2023MNRAS.524.2883N}. \citet{2025A&A...693A..33L} interpret this finding as being symptomatic of the lack of realism of a parametric description of the DM distribution in such a complex merging cluster.

In these previous comparisons, the image plane RMS, which quantifies how well lens models reproduce all image positions used in the reconstruction, has been employed for evaluating lens models. However, it does not directly assess predictions for images that were not used as constraints. \citet{2016ApJ...832...82J} showed that models with lower RMS values for the images used in the reconstruction can have poorer predictive performance when evaluated using the complete set of multiple images. The RMS of the fitted images can thus be considered a measure of reproduction accuracy, while the prediction of unused images provides a test of predictive accuracy. This distinction is relevant when evaluating whether allowing halo offsets improves a lens model.

We use the jackknife method to test whether offsets between BCGs and cluster-scale halos in the lens models are required by SL data themselves in this study. The method evaluates predictions for source systems excluded from the lens reconstruction and has been used to validate cluster lens models and investigate possible overfitting \citep{2025OJAp....8E.123N,2025PhRvD.112l3526L}. Following \citet{2025OJAp....8E.123N}, we assess model accuracy through its ability to predict the positions of multiple images not used as constraints. We compare models with fixed and free cluster-scale halo positions using the same excluded systems. A significant difference in predictive performance would provide evidence that the SL data prefer one of these positional assumptions.

We apply this test to three galaxy clusters: Abell 370 ($z=0.375$, hereafter A370), RX J1347.5-1145 ($z=0.451$, hereafter RXJ1347), and MACS J0416.1-2403 ($z=0.396$, hereafter MACSJ0416). Previous SL studies of these clusters have inferred offsets between the fitted cluster-scale halos and BCGs \citep[e.g.,][]{2014MNRAS.437.1858K,2023MNRAS.524.2883N,2025A&A...696A..15R}. We construct the lens models using the hybrid algorithm {\tt MrMARTIAN} \citep{2026ApJ...997...18C}, which extends the free-form approach of {\tt MARS} \citep{2022ApJ...931..127C,2023ApJ...951..140C} by combining analytic mass profiles with a regularized grid. The grid represents mass structure not captured by the analytic profiles, allowing the total mass distribution to adjust even when the cluster-scale halo centers are fixed. Since both the fixed and free models retain this flexibility, this framework allows us to test whether shifting the analytic halo centers provides an improvement in predictive accuracy when additional mass structure can be represented by the grid.

This paper is outlined as follows. Section~\ref{sec:data} presents the SL data, and Section~\ref{sec:jackknife} describes the jackknife method. The lens modeling framework and model configurations are presented in Section~\ref{sec:lens_modeling}. In Section~\ref{sec:results}, we present the jackknife results, total mass distributions, and fitted halo positions. Section~\ref{sec:discussion} discusses the interpretation of halo offsets and the effects of halo positional freedom and regularization on magnification, time-delay predictions, and critical curve positions around the Dragon Arc in A370. We summarize our conclusion in Section~\ref{sec:summary}.

We assume a flat $\Lambda$CDM cosmology with the matter density $\Omega_{\rm M}=1-\Omega_{\Lambda}=0.3$ and the dimensionless Hubble constant parameter $h=0.7$ unless stated otherwise. At the redshifts of A370 ($z=0.375$), RXJ1347 ($z=0.451$), and MACSJ0416 ($z=0.396$), the corresponding plate scales are $5.16, ~5.77$, and $5.34\ {\rm kpc ~ arcsec^{-1}}$, respectively.

\section{SL Data} \label{sec:data}
\begin{figure*}
\centering
\includegraphics[width=\textwidth]{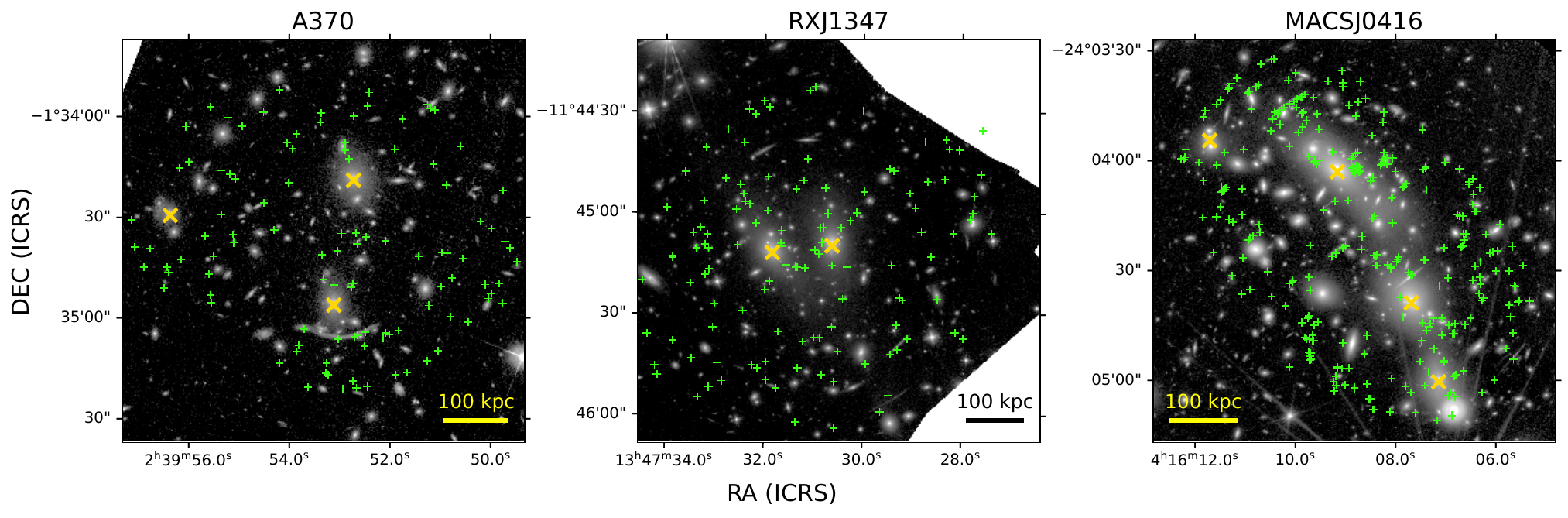} 
\caption{Multiple image distributions and initial locations of cluster-scale halos for lens reconstructions. The green crosses indicate the multiple image distributions while the yellow crosses mark the initial positions of cluster-scale halos. The mosaic images obtained from the Hubble Legacy Archive of A370 and RXJ1347 are created using the HST F814W and F160W filters, respectively. The mosaic image of MACSJ0416 is generated from the JWST F200W filter.} 
\label{fig:input_data}
\end{figure*}

Figure~\ref{fig:input_data} shows the multiple image distributions and initial positions of cluster-scale DM halos for the jackknife analysis of each target galaxy cluster. We use spectroscopically confirmed SL multiple images from previous studies. For A370, we adopt 105 multiple images from 35 source systems compiled by \citet{2023ApJ...951..140C} based on the catalogs of \citet{2019MNRAS.485.3738L} and \citet{2021MNRAS.506.6144G}. For RXJ1347, we use 125 multiple images from 38 source systems published by \citet{2021A&A...646A..83R}. For MACSJ0416, we use 303 multiple images from 111 source systems published by the CANUCS team\footnote{\url{https://niriss.github.io/lensing.html}} \citep{2025A&A...696A..15R}. For more details on these multiple image catalogs, we refer readers to the corresponding studies.

\section{Jackknife Method}\label{sec:jackknife}
The image plane RMS has been widely used in SL modeling to quantify how well a lens model reproduces the observed positions of the multiple images used as constraints. However, it does not assess the predictive power of the model for multiple images excluded from the reconstruction or newly added to the catalog. The image plane RMS can therefore be considered a measure of reproduction accuracy rather than predictive accuracy.
Following \citet{2025OJAp....8E.123N}, we assess the accuracy of a lens model in terms of its predictive power, defined as its ability to predict the positions of multiple images not used in the lens reconstruction. The jackknife method evaluates this predictive accuracy by excluding each source system in turn. 

Previous studies used the jackknife method to diagnose possible overfitting \citep{2025OJAp....8E.123N} and to validate cluster lens models used to measure $H_0$ from the time delays of SN Refsdal \citep{2025PhRvD.112l3526L}. In this study, on the other hand, we use the jackknife method to test whether allowing cluster-scale analytic halos to be offset from BCGs improves the prediction of excluded multiple image systems. This comparison assesses whether such offsets are required by SL constraints within the adopted lens modeling framework. Here, we briefly describe the jackknife procedure. For more details, we refer readers to \citet{2025OJAp....8E.123N} and \citet{2025PhRvD.112l3526L}.

For a SL multiple image catalog containing $N_{\rm sys}$ source systems, we first exclude all multiple images belonging to the $k$th source system and reconstruct the lens model using the remaining $N_{\rm sys}-1$ systems. We then calculate the predicted positions of the excluded multiple images in the image plane and compare them with their observed positions. This procedure is repeated by excluding each source system in turn. As a result, we obtain $N_{\rm sys}$ jackknife lens models and an equal number of system RMS values.
For the $k$th excluded source system, we compute the RMS between the observed and predicted image positions as
\begin{equation}
R_{{\rm sys},k}
=
\left[
\frac{1}{N_k}
\sum_{i=1}^{N_k}
\left|
\boldsymbol{\theta}^{\rm obs}_{i,k}
-
\boldsymbol{\theta}^{\rm pred}_{i,k}
\right|^2
\right]^{1/2},
\end{equation}
where $\boldsymbol{\theta}^{\rm obs}_{i,k}$ and $\boldsymbol{\theta}^{\rm pred}_{i,k}$ are the observed and predicted positions of the $i$th multiple image, respectively. $N_k$ indicates the number of multiple images in the $k$th source system.
After repeating the procedure for all source systems, we calculate the mean system RMS as
\begin{equation}
\overline{R}_{\rm sys}
=
\frac{1}{N_{\rm sys}}
\sum_{k=1}^{N_{\rm sys}}
R_{{\rm sys},k}.
\end{equation}
For each pair of lens models constructed under the two assumptions, we compare their $\overline{R}_{\rm sys}$ values using the same excluded systems. We also calculate the paired difference for each excluded system as
\begin{equation}
\Delta R_{{\rm sys},k}
=
R_{{\rm sys},k}^{\rm free}
-
R_{{\rm sys},k}^{\rm fixed},
\end{equation}
where free and fixed indicate models in which the cluster-scale halo positions are allowed to vary or fixed during the lens reconstruction, respectively.

A lower $\overline{R}_{\rm sys}$ indicates better overall predictive performance, while a systematic shift of $\Delta R_{{\rm sys},k}$ below or above zero indicates that the free or fixed model produces lower residuals across the excluded systems, respectively. These comparisons allow us to evaluate whether the offsets between BCGs and cluster-scale halos inferred in the lens models are required by SL data. We apply the jackknife framework for this comparison to the lens models described in Section~\ref{sec:lens_modeling}.

\section{Lens Modeling}\label{sec:lens_modeling}

\subsection{MrMARTIAN Framework}
Here, we use the hybrid lens modeling algorithm {\tt MrMARTIAN} \citep[][]{2026ApJ...997...18C}. {\tt MrMARTIAN} reconstructs the mass distribution by combining a regularized grid with analytic mass profiles. The total convergence and deflection are calculated by summing the contributions from both components. The grid allows positive and negative convergence values, enabling it to represent local mass structure above or below that described by the analytic profiles. This approach provides additional flexibility in reconstructing the mass distribution even when the cluster-scale halo centers are fixed, like other hybrid lens modeling algorithms  \citep[e.g.,][]{2014MNRAS.437.2642S, 2021MNRAS.506.2002B}.

{\tt MrMARTIAN} minimizes the following objective function:
\begin{equation}
f = \chi^2_{\rm SL} + r\mathcal{R},
\label{eq:lens_model_objective}
\end{equation}
where $\chi^2_{\rm SL}$ measures the agreement among the source positions inferred from the multiple images, $\mathcal{R}$ is the regularization term applied to the grid, and $r$ determines its relative weight.
We adopt an image plane positional uncertainty of $\sigma_{\rm img}=0\farcs4$ for all multiple images. 
We note that this value is a modeling assumption for comparing the fixed and free models under the same conditions, rather than an estimate of the actual positional uncertainty of each image.
The image plane covariance is $\mathbf{C}_{\rm img}=\sigma_{\rm img}^{2}\mathbf{I}$, where $\mathbf{I}$ is the identity matrix. We propagate this uncertainty into the source plane as
\begin{equation}
\mathbf{C}_{{\rm src},i,k} = \mathbf{A}_{i,k}\mathbf{C}_{\rm img}\mathbf{A}_{i,k}^{\rm T},
\label{eq:source_covariance}
\end{equation}
where $\mathbf{A}_{i,k}=\partial\boldsymbol{\beta}/\partial\boldsymbol{\theta}$ is the Jacobian of the lens mapping evaluated at the observed position of the $i$th image in the $k$th source system. The SL term is then
\begin{equation}
\chi^2_{\rm SL} = \sum_k\sum_{i=1}^{N_k}\left(\boldsymbol{\beta}_{i,k}-\overline{\boldsymbol{\beta}}_k\right)^{\rm T}\mathbf{C}_{{\rm src},i,k}^{-1}\left(\boldsymbol{\beta}_{i,k}-\overline{\boldsymbol{\beta}}_k\right),
\label{eq:source_chisq}
\end{equation}
where $\boldsymbol{\beta}_{i,k}=\boldsymbol{\theta}^{\rm obs}_{i,k}-\boldsymbol{\alpha}(\boldsymbol{\theta}^{\rm obs}_{i,k},z_k)$ is the source position inferred from that image, and $\overline{\boldsymbol{\beta}}_k$ is the covariance-weighted mean source position of the system. The deflection $\boldsymbol{\alpha}$ is evaluated at the spectroscopic source redshift $z_k$.
We note that the number of predicted multiple images is not explicitly constrained in our source plane optimization. We will discuss this aspect in Section~\ref{sec:results}.

For the grid, we use the maximum-entropy regularization implemented in {\tt MrMARTIAN}. This regularization accommodates both positive and negative grid convergence and suppresses spurious fluctuations. Its form is
\begin{equation}
\mathcal{R} = \sum\left[p_+ + p_- - \psi + \kappa_{\rm grid}\ln\left(\frac{\psi+\kappa_{\rm grid}}{2p_+}\right)\right],
\label{eq:grid_regularization}
\end{equation}
with
\begin{equation}
\psi = \left(\kappa_{\rm grid}^{2}+4p_+p_-\right)^{1/2},
\label{eq:grid_entropy_auxiliary}
\end{equation}
where $p_+$ and $p_-$ are the positive prior maps for the positive and negative components of the grid convergence, respectively (see \citealt{1998MNRAS.298..905H} for more details). We update the prior iteratively using the Gaussian smoothed grid from the previous minimization epoch. For more details, we refer readers to \citet{2026ApJ...997...18C}.

We employ $50\times50$ grids with a margin of 20 pixels. The modeled FOV is $120''\times120''$ for A370 and RXJ1347, and $110''\times110''$ for MACSJ0416. The margin accounts for the deflection contribution from mass outside the modeled FOV and reduces boundary effects in the convolution used to compute the grid deflection. We optimize the grid values and free analytic parameters until the objective function no longer decreases.

\subsection{Analytic Components}

The original application of {\tt MrMARTIAN} used truncated pseudoelliptical NFW profiles for its analytic component. In our analysis, we use pseudo-isothermal elliptical mass distributions (PIEMDs) for cluster-scale halos \citep{1993ApJ...417..450K} and dual pseudo-isothermal elliptical mass distributions (dPIEs) for cluster member galaxies \citep{2001astro.ph..2341K, 2007arXiv0710.5636E}.

For both profiles, we express the lensing strength through the angular Einstein radius parameter $\theta_{\rm E,\infty}$, adopting the source-independent normalization used for our mass maps \citep[see also][]{2025A&A...702A.157E}. Each component has its own center $(x_0,y_0)$, projected axis ratio $q$, and position angle $\phi$. Here $\phi$ is measured from the coordinate $x$ axis toward the major axis. Coordinates aligned with that axis are
\begin{align}
X &= (x-x_0)\cos\phi+(y-y_0)\sin\phi, \\
Y &= -(x-x_0)\sin\phi+(y-y_0)\cos\phi.
\end{align}
We define the elliptical radius in terms of the ellipticity $e$ as
\begin{equation}
R^{2} = \frac{X^{2}}{(1+e)^{2}}+\frac{Y^{2}}{(1-e)^{2}}.
\label{eq:profile_elliptical_radius}
\end{equation}
Using $e=(1-q)/(1+q)$, the same radius can also be expressed in terms of the axis ratio:
\begin{equation}
R^{2} = \left(\frac{1+q}{2}\right)^{2}\left(X^{2}+\frac{Y^{2}}{q^{2}}\right).
\label{eq:profile_elliptical_radius_axis_ratio}
\end{equation}

The PIEMD convergence profile is
\begin{equation}
\kappa_{\rm PIEMD}(x,y) = \frac{\theta_{\rm E,\infty}}{2\sqrt{R^{2}+r_{\rm core}^{2}}},
\label{eq:piemd_convergence}
\end{equation}
where $r_{\rm core}$ sets the angular size of the central core. The profile has a finite central convergence for $r_{\rm core}>0$ and approaches $\kappa\propto R^{-1}$ outside the core. No outer truncation is imposed on this component. Its six parameters are $(x_0,y_0,\theta_{\rm E,\infty},r_{\rm core},q,\phi)$. The central coordinates are fixed or allowed to vary according to the positional assumption being tested, while the remaining parameters are fitted. In addition, to describe each cluster member galaxy, we use the difference of two cored isothermal terms with a common center, axis ratio, and position angle:
\begin{equation}
\kappa_{\rm dPIE}(x,y) = \frac{\theta_{\rm E,\infty}}{2}\frac{r_{\rm cut}}{r_{\rm cut}-r_{\rm core}}\left[\frac{1}{\sqrt{R^{2}+r_{\rm core}^{2}}}-\frac{1}{\sqrt{R^{2}+r_{\rm cut}^{2}}}\right],
\label{eq:dpie_convergence}
\end{equation}
where $0< r_{\rm core}<r_{\rm cut}$ and $r_{\rm cut}$ is the angular truncation scale. Beyond the truncation scale, the convergence decreases as $R^{-3}$, giving a finite total projected mass. The seven parameters are $(x_0,y_0,\theta_{\rm E,\infty},r_{\rm core},r_{\rm cut},q,\phi)$. We fix $r_{\rm core}=0.001''$ for all dPIE profiles.

To decrease the number of free parameters, we adopt the luminosity scaling relation for cluster member galaxies:
\begin{equation}
\theta_{{\rm E},\infty,i} = \theta_{{\rm E},\infty,{\rm ref}}\left(\frac{L_i}{L_{\rm ref}}\right)^{2\alpha},\qquad r_{{\rm cut},i} = r_{{\rm cut},{\rm ref}}\left(\frac{L_i}{L_{\rm ref}}\right)^{\beta},
\label{eq:dpie_luminosity_scaling}
\end{equation}
where $L_i$ is the luminosity of the $i$th galaxy and $L_{\rm ref}$ is the reference luminosity. 
In cluster lens modeling, only the normalization parameters $\theta_{{\rm E},\infty,{\rm ref}}$ and $r_{{\rm cut},{\rm ref}}$ are fitted to describe cluster members that follow the luminosity scaling relations. No independent scatter around these relations is allowed for individual galaxies.
Here, we use HST/ACS F814W magnitudes for A370 and HST/WFC3 F160W magnitudes for RXJ1347 and MACSJ0416. In the case of A370 and RXJ1347, we adopt $\alpha=0.25$ and $\beta=0.5$. For MACSJ0416, we adopt $\alpha=0.3$ and $\beta=0.6$, following the scaling relations adopted by \citet{2023A&A...674A..79B}.
We include 128, 139, and 213 cluster member galaxies in A370, RXJ1347, and MACSJ0416, respectively. Galaxies treated individually are not constrained by the luminosity scaling relations. We fit the two BCGs individually in A370 and RXJ1347. For MACSJ0416, we fit galaxy ID 8971 and the foreground galaxy individually, following \citet{2023A&A...674A..79B} and \citet{2025A&A...696A..15R}. For these individually fitted galaxies, the positions, axis ratios, and position angles are fixed, while the Einstein radius parameters and cut radii are fitted.

\subsection{Model Configurations}

\begin{figure*}
\centering
\includegraphics[width=\textwidth]{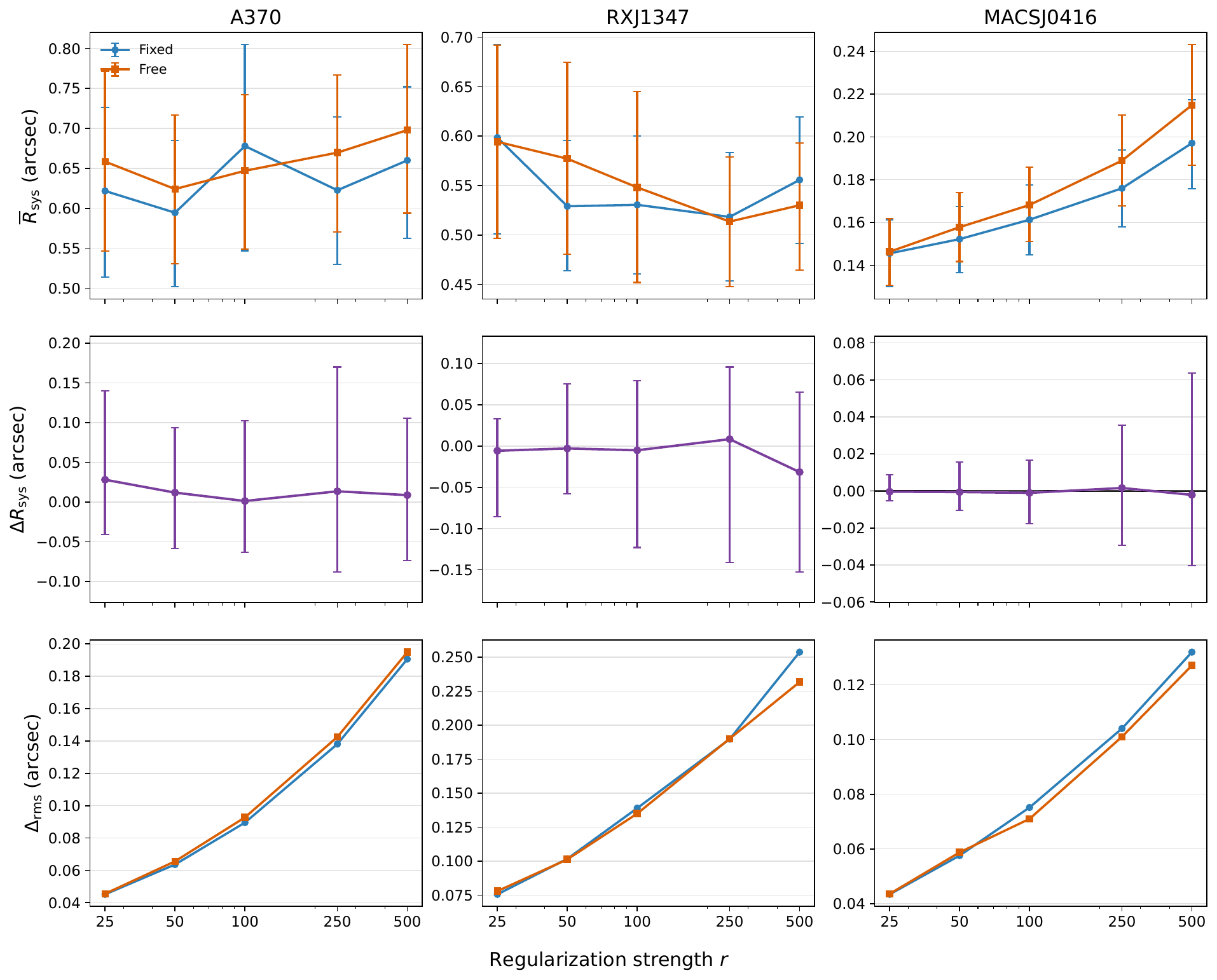}
\caption{Comparison of the jackknife predictive performance and image plane RMS of the fixed and free models across different regularization weights. In the upper and lower rows, blue and orange indicate the fixed and free models, respectively. The upper row shows the $\overline{R}_{\rm sys}$ values, with standard deviations obtained from 10,000 bootstrap realizations of the source systems. The middle row presents the medians and 16th to 84th percentile ranges of the paired differences $\Delta R_{{\rm sys}}$. The lower row shows the image plane RMS values, $\Delta_{\rm rms}$, for the models reconstructed using all multiple images.}
\label{fig_rms_mean}
\end{figure*}

We use three, two, and four cluster-scale PIEMD profiles for A370, RXJ1347, and MACSJ0416, respectively. Among the cluster-scale DM halos included in previous lens models \citep[e.g.,][]{2014MNRAS.437.1858K,2023MNRAS.524.2883N,2025A&A...696A..15R}, we use only those with optical counterparts. Their initial positions are indicated by the yellow crosses in Figure~\ref{fig:input_data}.

To test the positional freedom of cluster-scale halos, we construct fixed and free halo position models. In the model with fixed halo positions (hereafter fixed model), the PIEMD centers are fixed at the positions of the BCGs or bright cluster galaxies adopted in the previous lens models. In the model with free halo positions (hereafter free model), both center coordinates are allowed to vary within $[-15,15]$ arcseconds relative to their fixed values. All other PIEMD parameters are fitted in both cases.

In parametric SL modeling, positional uncertainties are often adjusted so that the reduced $\chi^2$ is close to unity. In our models, however, the number of free parameters exceeds the number of SL constraints. The effective number of free parameters can be estimated using the Fisher information matrix and the curvature of the regularization term \citep[e.g.,][]{Spiegelhalter_2002}, but this requires a more involved calculation. We keep the adopted positional uncertainty fixed and use the regularization weight to control the flexibility of the grid and limit overfitting. We employ five different regularization weights, $r=25$, 50, 100, 250, and 500, to examine the dependence on regularization. At each regularization weight, the fixed and free models use the same SL catalog, grid setup, galaxy components, and remaining parameter priors. For each cluster, we define ten model configurations by combining the two halo-position assumptions with the five regularization weights, and apply the jackknife procedure described in Section~\ref{sec:jackknife} to each configuration. After the jackknife test, we also reconstruct each configuration using all multiple images for the comparisons of total mass distributions and lensing predictions.

A potential concern is the dependence of our results on the adopted value of $\sigma_{\rm img}$. Since the source-plane covariance in Equation~\ref{eq:source_covariance} scales with $\sigma_{\rm img}^2$, Equation~\ref{eq:lens_model_objective} can be rewritten as 
\begin{equation}
\sigma_{\rm img}^2 f =
\sigma_{\rm img}^2 \chi_{\rm SL}^2
+ (r\sigma_{\rm img}^2)R.
\end{equation}
For a common $\sigma_{\rm img}$ applied to all multiple images, the dependence on $\sigma_{\rm img}$ in the first term on the right-hand side cancels out. Therefore, a change in the common positional uncertainty can be compensated by adjusting $r$ to keep $r\sigma_{\rm img}^2$ fixed. Our tests with different regularization weights also explore the dependence on this relative weighting. We further examine this relation in Section~\ref{subsec_mag_TD}.

\section{Result}\label{sec:results}
\subsection{Jackknife Predictive Performance}\label{subsec:jackknife_result}

Figure~\ref{fig_rms_mean} shows the $\overline{R}_{\rm sys}$ and $\Delta R_{{\rm sys}}$ distributions across different cluster-scale halo positional assumptions and regularization weights. For all three clusters, the $\overline{R}_{\rm sys}$ values of the fixed and free models are consistent within their uncertainties, while the median $\Delta R_{{\rm sys}}$ values remain close to zero and their 16th to 84th percentile ranges include zero in all cases.  
We also test whether systems with more predicted multiple images than observed affect our jackknife results. Some systems are found to have more predicted images than observed. To examine whether these additional images affect our results, we exclude systems for which the models predict at least one or at least two additional images compared to observations. We then recalculate $\overline{R}_{\rm sys}$ and $\Delta R_{\rm sys}$ using the remaining systems and also using only the excluded systems. Although $\overline{R}_{\rm sys}$ values change, we find no systematic improvement in predictive performance from allowing the cluster-scale halo positions to vary. Therefore, our main conclusion remains unchanged.

Since some systems have relatively larger $R_{\rm sys}$ values in some reconstructions, we examine the influence of systems with large $R_{\rm sys}$ values on the comparison between the fixed and free models. Excluding systems above the 95th and 90th percentiles lowers $\overline{R}_{\rm sys}$ but does not change our finding of comparable predictive accuracy between the fixed and free models.
Overall, the free models tend to show larger $\overline{R}_{\rm sys}$ values than the fixed models, although the differences remain within the uncertainties. These results provide no evidence that allowing offsets between BCGs and cluster-scale halos in the lens models improves predictive performance, indicating that such offsets are not required by the SL data to predict the excluded systems.

\begin{figure*}
\centering
\includegraphics[width=\textwidth]{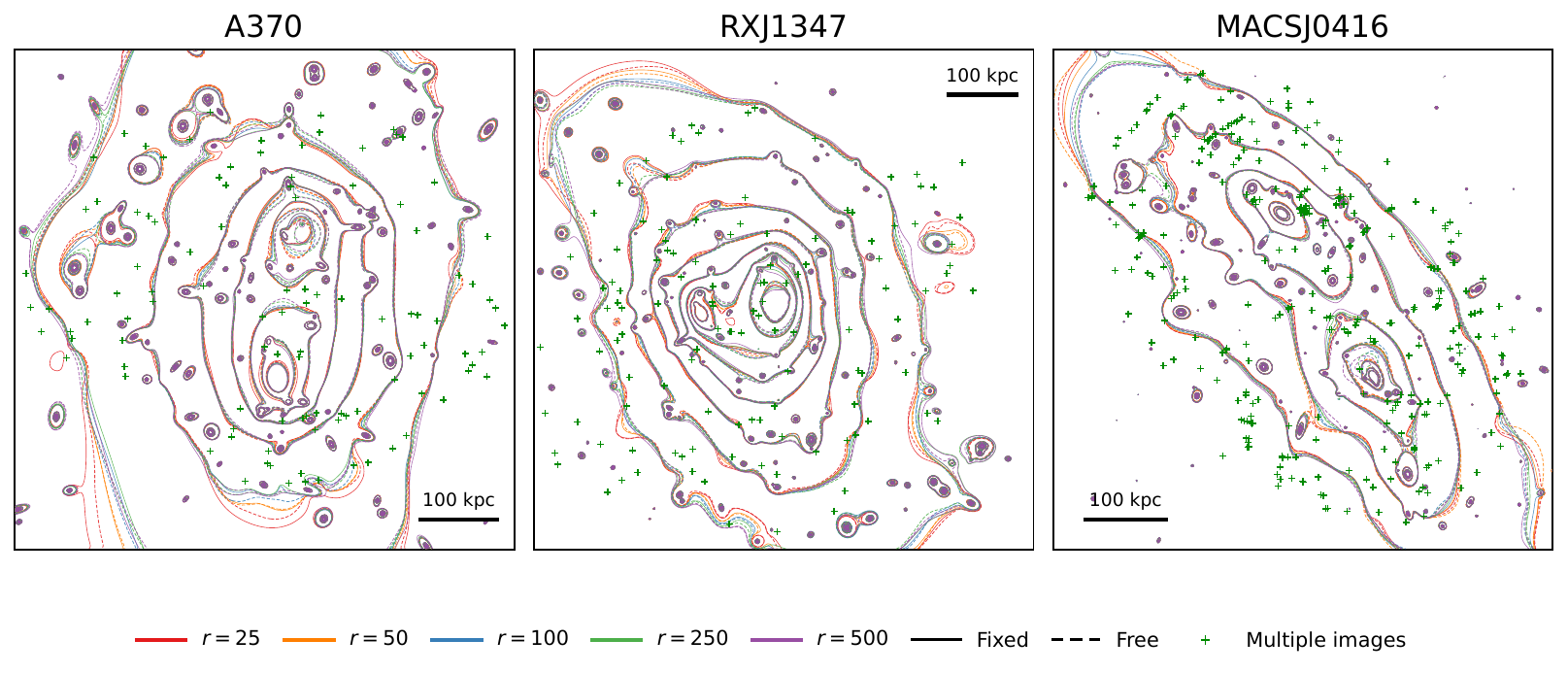} 
\caption{Mass contours of the fixed and free models across different regularization weights. Colors indicate the regularization weights, while solid and dashed lines represent the fixed and free models, respectively. The contours range from $\kappa=0.6$ to $2.4$ in intervals of $0.3$, with an additional $\kappa=3.0$ for RXJ1347. Green crosses mark the observed positions of the multiple images used as lensing constraints.} 
\label{fig:mass_contours}
\end{figure*}

\begin{figure}
\centering
\includegraphics[width = 0.775\columnwidth]{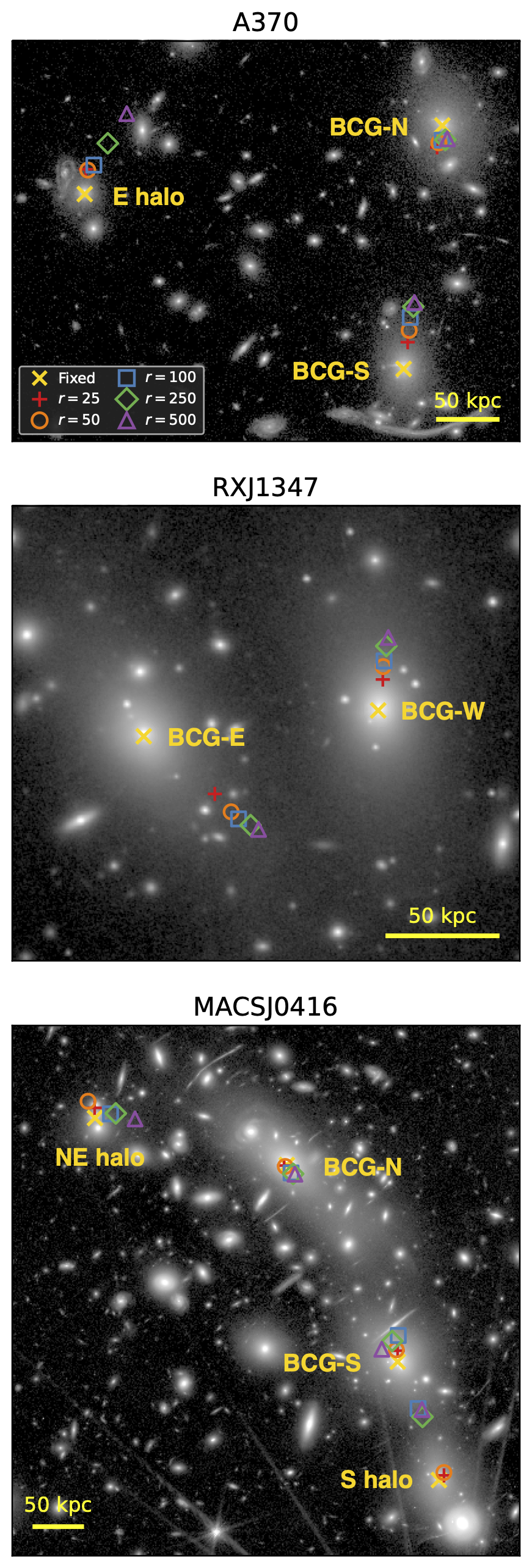}
\caption{Fitted positions of the cluster-scale PIEMD profiles in the free models. Yellow crosses indicate the fixed positions, while the colored markers show the fitted positions at different regularization weights.}
\label{fig:reg_halo_position}
\end{figure}

\begin{table*}
\caption{Projected offsets of the cluster-scale PIEMD centers from the optical positions adopted in the fixed models. Values are given in kpc for the free models reconstructed using all multiple images at each regularization weight. The halo labels correspond to those in Figure~\ref{fig:reg_halo_position}.}
\label{tab:halo_offsets}
\centering
\begin{tabular}{c|rrr|rr|rrrr}
\hline\hline
    & \multicolumn{3}{c|}{A370}
    & \multicolumn{2}{c|}{RXJ1347}
    & \multicolumn{4}{c}{MACSJ0416} \\
\cline{2-4}\cline{5-6}\cline{7-10}
$r$ & E halo & BCG-N & BCG-S
 & BCG-E & BCG-W
 & NE halo & BCG-N & BCG-S & S halo \\
\hline
25  & 19.66 & 17.51 & 21.89 & 40.14 & 13.73
    &  9.88 &  2.69 & 10.64 &  6.71 \\
50  & 19.08 & 13.97 & 30.93 & 50.64 & 19.35
    & 17.64 &  2.06 & 11.61 &  8.76 \\
100 & 24.29 & 11.63 & 40.92 & 55.15 & 21.86
    & 15.68 &  8.75 & 25.46 & 72.40 \\
250 & 44.44 & 12.20 & 50.38 & 60.96 & 28.46
    & 20.70 & 10.57 & 20.74 & 63.39 \\
500 & 85.52 & 11.77 & 52.53 & 64.79 & 32.31
    & 38.95 & 11.92 & 19.60 & 70.39 \\
\hline
\end{tabular}
\end{table*}

We also find no clear dependence of $\overline{R}_{\rm sys}$ on the regularization weight for A370 or RXJ1347. For MACSJ0416, however, $\overline{R}_{\rm sys}$ increases with increasing regularization weight in both the fixed and free models, indicating that their overall predictive performance decreases as the regularization weight increases. We suspect that this trend may be related to the much larger number of multiple images in MACSJ0416 (303) than in A370 (105) and RXJ1347 (125). 
However, the image plane RMS values ($\Delta_{\rm rms}$) increase with $r$ in both the fixed and free models for all three clusters. This trend may reflect the reduced flexibility of the grid under stronger regularization. Despite this increase in $\Delta_{\rm rms}$, $\overline{R}_{\rm sys}$ remains broadly similar within the bootstrap uncertainties for A370 and RXJ1347. These results show that a lower image plane RMS does not necessarily imply higher predictive accuracy, consistent with previous studies \citep[e.g.,][]{2016ApJ...832...82J,2025OJAp....8E.123N}.

\subsection{Mass Distribution}
Figure~\ref{fig:mass_contours} compares the total mass distributions reconstructed using all multiple images for the fixed and free models across the five regularization weights. For all three clusters, the overall shapes of the mass contours remain broadly similar regardless of the assumptions for the cluster-scale halo positions or the regularization weights. Within the SL region, the contours generally show close agreement, although local differences between the fixed and free models are found around the northern BCG of A370 and the southern BCG of MACSJ0416. Additional differences are seen in the outskirts, which may arise from the lack of multiple image constraints in these regions. This overall agreement, combined with the variation in the fitted positions of the cluster-scale PIEMD profiles, shows that the internal mass decomposition is not unique.

Previous studies also illustrate this dependence on the adopted mass representation. In MACSJ0416, \citet{2025A&A...703A..10L} obtained similar total mass distributions with three and four cluster-scale halos. Our fourth halo is associated with a southern bright galaxy included in their galaxy component, and the total projected mass within $10\arcsec$ of this galaxy is similar in the two studies. The additional flexibility of our grid may also explain why we can reproduce the SL constraints in A370 with three fixed halos, whereas the parametric models of \citet{2025A&A...693A..33L} required an additional dark clump and a displaced halo.

Figure~\ref{fig:reg_halo_position} shows the fitted positions of the cluster-scale PIEMD profiles in the free models, and Table~\ref{tab:halo_offsets} lists the projected offsets from the fixed positions. Despite the broadly similar total mass distributions shown in Figure~\ref{fig:mass_contours}, the magnitudes of the offsets from the BCG positions vary with regularization, while their directions remain broadly similar. The degree of this variation differs among the individual halos. In RXJ1347, for example, the fitted positions of both PIEMD profiles remain displaced from the BCG positions across all regularization weights, although the total mass contours remain similar to those of the fixed models. 
Within the {\tt MrMARTIAN} framework, these SL modeling results indicate that the mass decomposition among the regularized grid, cluster-scale PIEMD profiles, and galaxy-scale dPIE profiles is not uniquely determined.
Together with the jackknife results, this shows that lens models with different internal mass decompositions can produce similar total mass distributions while achieving comparable accuracy in predicting the positions of multiple images excluded from the reconstruction.

\section{Discussion}\label{sec:discussion}
\subsection{Interpretation of Halo-position Freedom}
Our results show no significant difference in predictive performance between models with fixed and free cluster-scale halo positions across all tested regularization weights. SL modeling reproduces the observed multiple image positions using the total deflection field, allowing different internal mass decompositions to produce similar lensing observables. Our results illustrate this non-uniqueness within the {\tt MrMARTIAN} framework, where different contributions from the regularized grid and cluster-scale PIEMD profiles reproduce broadly similar total mass distributions.

An inferred halo offset may reflect a genuine physical displacement, biases in the lens reconstruction, or both. These offsets can depend on how the DM center is defined and the spatial scale used to measure it \citep{2024OJAp....7E..65R}. LOS structure can also shift the relative positions of multiple images \citep{2026arXiv260530433R}, potentially affecting the fitted halo positions. Also, physical offsets may arise during cluster mergers. SIDM merger simulations can produce offsets of a few tens of kpc, depending on the cross section and merger phase \citep{2024OJAp....7E..65R, 2024MNRAS.529.2032S}. A370 has been modeled as a major merger after its second core passage \citep{2020ApJ...900..151M}, while simulations of equal mass mergers with self-interacting DM show that offsets evolve with merger phase and that BCGs can continue to oscillate around the DM center after the halos coalesce \citep{2017MNRAS.469.1414K}.

However, before considering the physical origin of offsets, it is necessary to establish whether a displaced cluster-scale halo is required in the lens model reconstruction. Our jackknife tests show that models with and without fitted halo offsets provide comparable predictions for excluded image systems, even in MACSJ0416 with more than 300 multiple images. Within the adopted framework, a fitted offset can be regarded as an allowed model solution rather than a feature uniquely established by the SL data. Another form of non-uniqueness is shown by \citet{2026arXiv260927147W}, who constructed different total mass distributions while preserving the deflection angles at the observed SL image positions. Given the dependence of fitted offsets on different mass decompositions, there are limitations to inferences about cluster dynamics or DM physics based on them alone.

\subsection{Impacts on Lensing Predictions}
\subsubsection{Magnification and Time Delay}\label{subsec_mag_TD}
\begin{figure*}
\centering
\includegraphics[width = \textwidth]{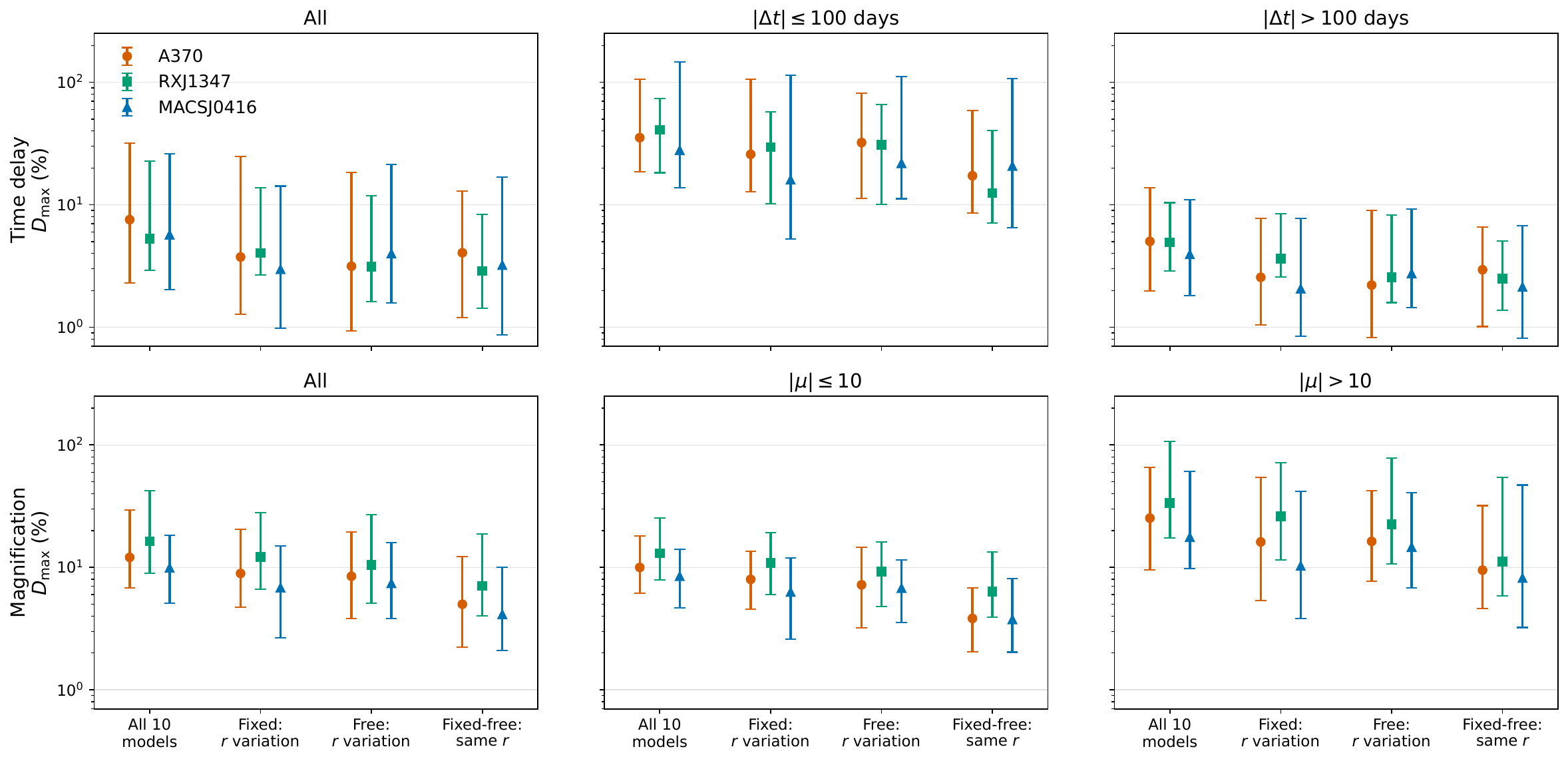}
\caption{Maximum relative differences in time delay (top row) and magnification (bottom row) within the tested model sets, as defined in Equation~\ref{eq:dif_max}. The left column shows the full sample. The middle and right columns separate the sample at $|\Delta t|=100$ days for time delay and $|\mu|=10$ for magnification, with values at or below the thresholds in the middle column and above them in the right column. This classification uses the median absolute predicted value across the ten models for each image pair or individual image.  Within each panel, the comparisons from left to right include all ten models, variation in $r$ within the fixed models, variation in $r$ within the free models, and fixed-free differences at the same $r$. For the last comparison, the maximum is taken among the five regularization weights. Circles, squares, and triangles represent A370, RXJ1347, and MACSJ0416, respectively. The markers and error bars indicate the medians and 16th to 84th percentile ranges of $D_{\max}$ across image pairs for time delay and individual images for magnification.}
\label{fig:TD_magnification_predictions}
\end{figure*}

\begin{figure*}
\centering
\includegraphics[width=\textwidth]{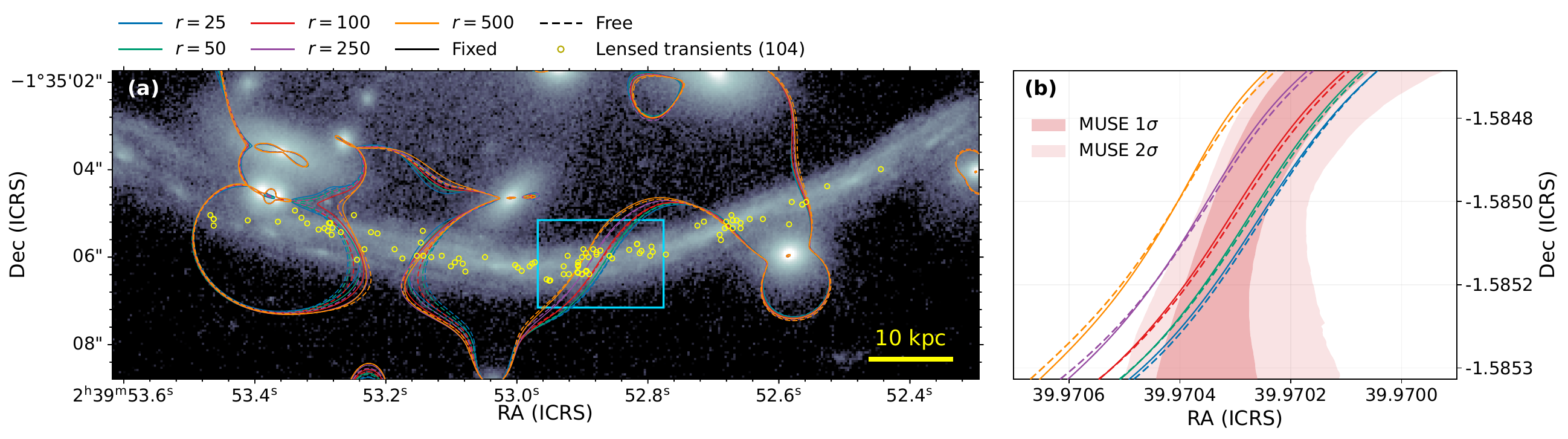} 
\caption{Critical curves predicted at the source redshift of the Dragon Arc ($z=0.725$) in A370. The colored curves indicate regularization weights $r=25$, 50, 100, 250, and 500. The solid and dashed lines represent the fixed and free models, respectively. Yellow open circles mark the positions of the 104 lensed transients reported in \citet{2026arXiv260422702P}. The cyan rectangle in panel (a) marks the region shown in panel (b). (b) Comparison between our predicted critical curves and the MUSE kinematic constraint from \citet{2026ApJ..1006...64Z}. The dark and light red shaded regions indicate the pointwise $1\sigma$ and $2\sigma$ posterior intervals, respectively.
} 
\label{fig:mag_dragon_arc}
\end{figure*}

To quantify the maximum differences in magnification and time delay across the tested model sets, we define the maximum relative difference as
\begin{equation}
D_{\max,i}
=
\max_{(m,n)\in P}
\left[
\frac{2\left|Q_i^{(m)}-Q_i^{(n)}\right|}
{\left|Q_i^{(m)}\right|+\left|Q_i^{(n)}\right|}
\right]\times 100,
\label{eq:dif_max}
\end{equation}
where $Q_i^{(m)}$ denotes the signed magnification of image $i$ or the signed time delay of image pair $i$ predicted by model $m$, and $P$ specifies the model pairs being compared. 
Magnifications and time delays are calculated at the image positions predicted by each model, with time delay measured relative to the first image of each system. 
The symmetric normalization treats both models equally without adopting either as a reference.
We note that $D_{\max}$ is a metric to evaluate the differences among the tested models, not the uncertainties of the predicted values. In real observations, positional errors of multiple images of point-like sources, such as SNe or QSOs, can be smaller than those of extended images. These differences in positional errors can affect the uncertainties in the predicted values and should be considered for individual sources.

In Figure~\ref{fig:TD_magnification_predictions}, we summarize the resulting $D_{\max}$ distributions for the full sample and for subsamples divided by the predicted magnification or time delay. We separate the samples at $|\mu|=10$ and $|\Delta t|=100$ days, using the median absolute predicted value across the ten models for each image or image pair. For each sample, we consider four comparisons: 1) all pairs among the ten models with $r=25$, 50, 100, 250, and 500, 2) all pairs among the five fixed models, 3) all pairs among the five free models, and 4) the five fixed-free pairs with the same $r$. For the last comparison, we retain the largest relative difference across the five regularization weights for each image or image pair.
 
For the full sample, the median $D_{\max}$ values among all ten models are $\sim10\%$ for both magnification and time delay. For magnification, the medians of the maximum fixed-free differences across the tested regularization weights are lower than those obtained by varying $r$. Within the {\tt MrMARTIAN} framework, these comparisons indicate that magnification predictions are more sensitive to regularization than to halo positional freedom over the tested range of $r$. For time delay, however, varying $r$ and changing the halo positional assumption yield comparable median $D_{\max}$ values of $\sim3$–$4\%$. 

However, the subsample comparisons reveal larger relative differences for highly magnified images and image pairs with short time delays, although the median fixed-free differences remain lower than those obtained by varying $r$ in both subsamples, preserving the trend found for the full sample. The magnification subsamples show that relative differences are larger for images with $|\mu|>10$ than for those with $|\mu|\leq10$. Individual images can show larger differences than the medians, as reflected in the upper percentile ranges. Large variations in magnification between parametric models were also reported \citep[e.g.,][]{2025A&A...693A..33L}. For time delay, the relative differences are clearly larger for image pairs with $|\Delta t|\leq100$ days than for those with longer delays across all four comparisons. The median maximum fixed-free differences are $\sim10$-$20\%$ for the shorter delay subsample, compared with $\sim2$-$3\%$ for the longer delay image pairs.

We also examine the dependence of magnification and time-delay predictions on the adopted positional uncertainty. For all three clusters, we reconstruct additional fixed and free models using $(r,\sigma_{\rm img})=(100,0\farcs2)$ and $(100/9,0\farcs6)$, corresponding to the reference configuration of $(25,0\farcs4)$ with the same $r\sigma_{\rm img}^2$. The resulting magnification and time-delay predictions generally agree with those of the reference models, confirming that a common change in $\sigma_{\rm img}$ can be compensated by adjusting the regularization weight.

The comparison in Figure~\ref{fig:TD_magnification_predictions} show that lens models with similar total mass distributions can yield different magnification and time-delay predictions. This model dependence highlights the need to account for systematic uncertainties arising from different modeling assumptions when using cluster lens models to study highly magnified sources or to infer $H_0$ from time delays. 
For example, some cluster-scale time delay cosmography studies have used delays of $\sim100$ days or less to infer $H_0$ \citep[e.g.,][]{2025ApJ...979...13P,2026A&A...708A.291S}, which are comparable to the delays in the subsample for which our model comparisons show larger $D_{\max}$ values. Differences in magnification predictions can affect the intrinsic properties inferred for highly magnified sources at high redshift \citep[e.g.,][]{2022ApJ...927...81B,2026OJAp....966816A}. We note that our comparisons are based on only three clusters, and a larger sample is needed to assess whether these trends hold more generally.

\subsubsection{Critical Curves}

In addition to magnification, halo positional freedom and regularization can affect the locations of critical curves. These locations are important for assessing whether compact sources at high redshift are individual lensed stars or star clusters, because the inferred magnification and intrinsic source size depend on their proximity to the critical curve \citep[e.g.,][]{2025ApJ...993..226S, 2025ApJ...988L..76P}. The spatial distribution of lensed transients relative to critical curves has been used to investigate the nature of DM \citep[e.g.,][]{2018PhRvD..97b3518O,2018NatAs...2..334K,2019A&A...625A..84D,2025NatAs...9..428F,2025A&A...699A.299M,2025A&A...704A.309L,2026arXiv260422702P}. Here, we examine the critical curves around the Dragon Arc at $z=0.725$ in A370.

Figure~\ref{fig:mag_dragon_arc} shows our predicted critical curves together with the positions of the reported stellar transients \citep{2025NatAs...9..428F, 2026arXiv260422702P}. The fixed and free models produce broadly similar critical curves at the same regularization weight, whereas the curves shift with changing regularization. 
The critical curves at $r=25,~50$, and 100 show good agreement in position and orientation with the local critical curve independently inferred from MUSE IFU kinematics by \citet{2026ApJ..1006...64Z}, which was not used as a constraint in our lens reconstruction. Our critical curves pass through all groups of reported stellar transients, consistent with the expected concentration of microlensing events near critical curves.

These comparisons show that critical curve positions depend on the regularization weight even within the same modeling framework. Such shifts change the magnification inferred for a compact source and can alter the parity assigned to a lensed transient. Given this model dependence, there are limitations to physical interpretations based solely on critical curve positions predicted by cluster-scale lens models, indicating the need to incorporate constraints beyond multiple image positions, such as kinematic measurements or the extended surface brightness of lensed images \citep[e.g.,][]{2019A&A...631A.130B, 2024ApJ...976..110A}.

\subsection{Regularization Selection for {\tt MrMARTIAN}}
The jackknife method can also provide a useful metric for the regularization weight selection in {\tt MrMARTIAN}. If the weight is too small, the model may overfit the image positions used as constraints, increasing the residuals of the excluded systems. If the weight is too large, excessive smoothing of the grid may suppress mass structure needed to predict the excluded images, again increasing their residuals. Comparing $\overline{R}_{\rm sys}$ across regularization weights can help identify a balance between these effects.
For example, in A370, both the fixed and free models have their lowest $\overline{R}_{\rm sys}$ values at $r=50$, and these models also show the closest agreement with the local critical curve constraint from MUSE kinematics based on the metric used in \citet{2026ApJ..1006...64Z} (see Figure~\ref{fig:mag_dragon_arc}). Although the differences in $\overline{R}_{\rm sys}$ remain within the bootstrap uncertainties, this correspondence suggests that jackknife tests may help to select the regularization weight within the {\tt MrMARTIAN} framework. 

However, individual source systems have different influences on the jackknife results, as the lens reconstruction responds nonlinearly to changes in the multiple image constraints.
For example, excluding a system that provides the only constraints in a particular region can leave the local mass distribution poorly constrained. Predicted image positions near a critical curve can also be sensitive to small changes in the lens model. A few such systems may strongly affect $\overline{R}_{\rm sys}$ and complicate regularization selection. 
Despite these limitations, A370 and RXJ1347 show a minimum in $\overline{R}_{\rm sys}$ at an intermediate regularization weight within the tested range, with larger values at both weaker and stronger regularization. This suggests that the jackknife may provide a practical approach to selecting the regularization strength, although further tests with a larger sample of clusters and simulated data are needed to assess its robustness. Nevertheless, the jackknife method remains useful for evaluating the preference between lens models constructed under different assumptions.

\section{Summary} \label{sec:summary}
We investigate whether offsets between BCGs and cluster-scale halos inferred from SL modeling are required by the SL data. We use jackknife tests to compare the predictive accuracy of lens models for A370, RXJ1347, and MACSJ0416 by excluding each source system in turn. We construct the models using {\tt MrMARTIAN}, which combines analytic mass profiles with a regularized grid. The grid can represent mass structure not captured by the analytic components, even when the halo centers are fixed. This allows us to test whether varying the halo positions improves predictive accuracy. We compare fixed and free halo position models with different regularization weights.

The $\overline{R}_{\rm sys}$ values of the fixed and free models agree within their bootstrap uncertainties, while the medians of the $\Delta R_{{\rm sys}}$ remain close to zero. Across all three clusters and all tested regularization weights, these results indicate comparable predictive accuracy and provide no evidence that the SL data prefer models with fitted halo offsets. 
The total mass distributions remain broadly similar, although the magnitudes of the fitted halo offsets vary with regularization. These findings illustrate the non-uniqueness of the mass decomposition between the regularized grid and cluster-scale halo profiles. Given this dependence on the mass decomposition, there are limitations to inferences about cluster dynamics or DM physics based on fitted offsets.

In addition, we evaluate the effects of halo positional freedom and regularization on magnification and time-delay predictions. For the full sample, the median maximum fixed-free differences across the tested regularization weights remain below $10\%$ for magnification and $5\%$ for time delay. However, the subsample comparisons show larger relative differences for shorter time delays ($|\Delta t|\leq100$ days) and higher magnifications ($|\mu|>10$). Halo positional freedom and regularization can also affect critical curve positions. We examine these effects around the Dragon Arc in A370. Although all our models produce critical curves that pass through all reported stellar transient groups as expected, the curve positions vary with regularization.

In this study, we examine three galaxy clusters for which SL models infer large offsets between BCGs and cluster-scale halos. Our results show that allowing such offsets does not significantly improve the accuracy of predictions for excluded multiple image systems, indicating that halo positional freedom is not required by the SL data within the adopted {\tt MrMARTIAN} framework. This comparison provides a test of the predictive support for fitted halo offsets before they are used to infer cluster dynamics or DM physics. Our results also emphasize the need to account for systematic uncertainties arising from different modeling assumptions in magnification and time-delay predictions.

\begin{acknowledgements}
We express our gratitude to Masamune Oguri and Keiichi Umetsu for helpful comments and suggestions. We also thank Liang Dai and Ruwen Zhou for sharing their critical curve constraints derived from MUSE kinematics. SC acknowledges this research was supported through the EACOA Fellowship awarded by the East Asia Core Observatories Association and supported by Basic Science Research Program through the NRF funded by the Ministry of Education (No. RS-2024-00413036).
ML acknowledges the Centre National de la Recherche Scientifique (CNRS) and the Centre National des Etudes Spatiale (CNES) for support.
MJJ acknowledges support for the current research from the National Research Foundation (NRF) of Korea under the programs 2022R1A2C1003130 and RS-2026-25607498.
This work is based on observations made with the NASA/ESA Hubble Space Telescope and obtained from the Hubble Legacy Archive, which is a collaboration between the Space Telescope Science Institute (STScI/NASA), the Space Telescope European Coordinating Facility (ST-ECF/ESAC/ESA), and the Canadian Astronomy Data Centre (CADC/NRC/CSA).
This work is also based on observations made with the NASA/ESA/CSA James Webb Space Telescope and obtained from the Mikulski Archive for Space Telescopes (MAST) at the Space Telescope Science Institute (STScI). The specific observations used in this work can be accessed via DOI: 10.17909/v0rx-d363.
\end{acknowledgements}

\bibliographystyle{aa}
\bibliography{SL_jackknife}

\end{document}